\documentstyle[11pt,newpasp,twoside,epsf]{article}
\def\edcomment#1{\iffalse\marginpar{\raggedright\sl#1\/}\else\relax\fi}
\reversemarginpar

\begin{document}
\title{The Next Generation Sky Survey and the Quest for Cooler Brown Dwarfs}
 \author{J. Davy Kirkpatrick}
\affil{Infrared Processing and Analysis Center, California Institute of
Technology, Pasadena, CA 91125, USA}

\begin{abstract}
The Next Generation Sky Survey (NGSS) is a proposed NASA MIDEX mission to 
map the entire sky in four infrared bandpasses -- 3.5, 4.7, 12, and 23 
$\mu$m. The seven-month mission will use a 50-cm telescope and 
four-channel imager to 
survey the sky from a circular orbit above the Earth. Expected 
sensitivities will be half a million times that of COBE/DIRBE at 3.5 
and 4.7 $\mu$m and a thousand times that of IRAS at 12 and 23 $\mu$m. 
NGSS will be particularly sensitive to brown dwarfs cooler 
than those presently known. Deep absorption in the methane 
fundamental band at 3.3 $\mu$m and a predicted 5-$\mu$m overluminosity will 
produce uniquely red 3.5-to-4.7 $\mu$m colors for such objects. 
For a limiting volume of 
25 pc, NGSS will completely inventory the Solar Neighborhood for 
brown dwarfs as cool as Gl 229B. At 10 pc, the census will be 
complete to 500 K. Assuming a field mass function with $\alpha = 1$, 
there could be one or more brown dwarfs warmer than 150 K lying
closer to the Sun than Proxima Centauri and detectable primarily 
at NGSS wavelengths. NGSS will enable estimates of the brown dwarf mass 
and luminosity functions to very cool temperatures and will provide both 
astrometric references and science targets for NGST.
\end{abstract}

\section{Introduction}

Large-area surveys such as the Two Micron All Sky Survey (2MASS; see other
contribution by Kirkpatrick) and the Sloan Digital Sky Survey (SDSS; see
contribution by Covey) have revolutionized the field of brown dwarf science.
However, these surveys are able to probe only to temperatures near 1000 K at
a distance of 10 pc, which leaves the realm of cooler brown dwarfs almost
entirely unexplored. Fortunately, the Next Generation Sky Survey (NGSS) 
has been proposed in a large part to characterize cooler brown dwarfs in 
the Solar Neighborhood. In \S2 the
NGSS project will be briefly described. Its role in the future of
brown dwarf studies is highlighted in \S3, and the current status
of the mission is given in \S4.

\section{An Overview of the NGSS Mission}

NGSS is a proposed seven-month mission to
survey the entire sky between 3.5 and 23 $\mu$m. The 50-cm
telescope flies is a circular orbit 500 km above the Earth. 
It is equipped with a four-channel imager comprised of HgCdTe and
Si:As 1024{$\times$}1024 arrays to survey the sky at the four wavelengths
given in Table 1.
The resolution of the telescope plus re-imaging optics is $\sim$5\arcsec\
except at 12 and 23 {$\mu$}m where it is near
the diffraction limit of 6\arcsec\ and 12{\arcsec}, respectively. The actual
field of view of the detectors is 38\arcmin{$\times$}38{\arcmin}. 

\begin{table}
\begin{center}
\caption{Wavelengths and Sensitivities Probed by NGSS}
\begin{tabular}{cccc}
\tableline
Wavelength & Note & Sensitivity& Sensitivity\\
($\mu$m)   &      & ($\mu$Jy)  & (mag)\\
\tableline
3.5 & $\sim$L-band & $\sim$24 &  $\sim$17.6\\
4.7 & $\sim$M-band & $\sim$30 &  $\sim$16.8\\
12  & $\sim$N-band & $\sim$131&  $\sim$13.3\\
23  &   ...        & $\sim$414&  $\sim$10.8\\
\tableline
\tableline
\end{tabular}
\end{center}
\end{table}

NGSS will cover the sky from a sun-synchronous polar orbit with a 6 am/6 pm 
ascending node like IRAS. As it circles the earth, the telescope performs
one revolution per orbit so that it is always pointing away from the earth.
A scanning mirror freezes a 38\arcmin{$\times$}38{\arcmin} field-of-view on 
the 
detectors every 8.8 seconds for a total integration time per frame of 6.6 sec.
The rate of reset for the scanning mirror is chosen 
so that there is a small overlap between consecutive frames,
and a great circle perpendicular to the Earth-Sun line is scanned every orbit.
There is a large overlap from one orbit to the next with the maximal overlap
occurring at the ecliptic poles. A cartoon of the scanning procedure is 
shown in Figure 1. 

\begin{figure}
\plotone{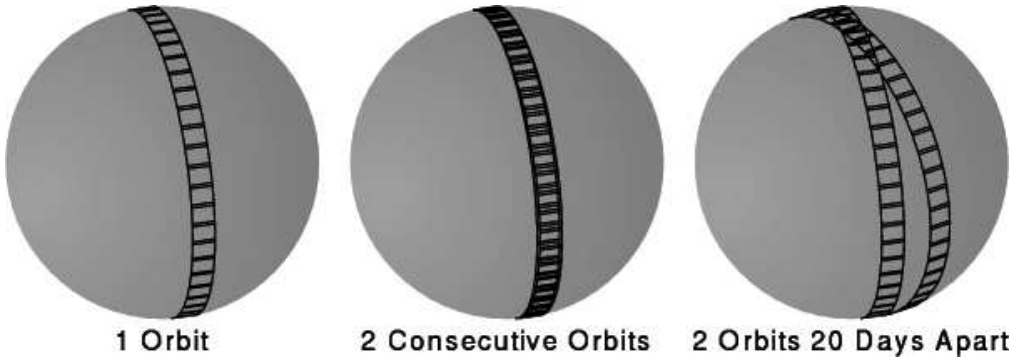}
\caption{Examples of the orbital scanning pattern. For clarity, the size of
the frames has been exaggerated and the number of frames per orbit reduced.}
\end{figure}

In August, 1998, NGSS was first proposed to NASA as a mission in the
Medium-class Explorer (MIDEX) program. Although selected as one of five
missions for Phase-A study in January, 1999, it was not selected as one
of the winners in the final round in October, 1999. NGSS was re-proposed
as a MIDEX mission in October, 2001, and was one of four missions selected
for Phase-A study in April, 2002.

\section{The NGSS Contribution to Brown Dwarf Science}

Approximate sensitivities (away from the ecliptic plane) 
are listed in Table 1. In the
two short wavelength bands, NGSS is 500,000 times deeper than COBE/DIRBE.
At its two long wavelength bands, it is 1,000 times deeper than the short
wavelength bands of IRAS. For the purposes of brown dwarf science, it is
the 3.5- and 4.7-$\mu$m bands that are the most useful. Here
NGSS is exploring wavelengths intermediate
between those of 2MASS and IRAS, it probes them deeply, and it 
covers the entire
sky. 

Figure 2 shows a schematic diagram of currently recognized brown dwarf
types. Thanks to large-area surveys such as 2MASS and SDSS, a large number of
field late-M, L, and T dwarfs are now recognized. 
According to the models of Burrows et al.\ (1997), for an age of 1 Gyr an
M8 dwarf would lie near the stellar/substellar border of 75 M$_{Jup}$, a
mid-L dwarf would have a mass of $\sim$65 M$_{Jup}$, and a mid-T would have
a mass of $\sim$30 M$_{Jup}$. The models predict effective
temperatures of 2200 K, 1700 K, and 1200 K for these same objects. The 2MASS
and SDSS surveys are sensitive to mid-T dwarfs only out to about 10 pc, and
the coolest object so far discovered by either survey has an effective 
temperature of only $\sim$750 K (Gliese 570D at $d = 5.9$ pc; Burgasser et al.\ 2000). 

\begin{figure}
\plotone{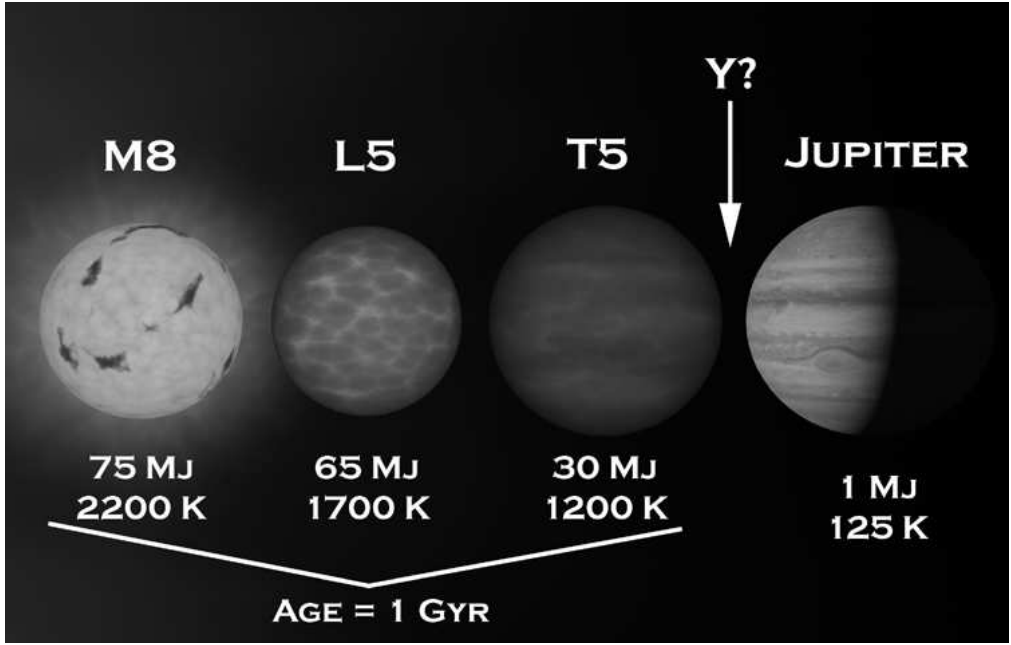}
\caption{An artist's rendition comparing three flavors of brown
dwarfs with Jupiter. All are plotted on the same size
scale and the three brown
dwarfs are chosen to have ages of 1 Gyr. On the far left 
is shown an M8 dwarf at the stellar/substellar border followed by a mid-L
dwarf and a mid-T dwarf. Temperatures and masses, derived from theory, are
listed on the figure. Note the wide gap in mass and temperature between
the mid-T dwarf and the planet Jupiter. It is within this gap that NGSS will
have its greatest impact for brown dwarf research, uncovering even cooler
objects (which we might call ``Y dwarfs'' if a new feature appears that is not
characteristic of T dwarfs; e.g., the
reshaping of the spectrum when H$_2$O clouds form below $T_{eff} \approx 
400$K) and revealing
just how prevalent brown dwarfs are in the solar vicinity.}
\end{figure}

The peak emission of even colder brown dwarfs lies squarely in the wavelength
regime in which NGSS operates. If such objects radiated like blackbodies,
a 500-K brown dwarf would have its peak emission at 5.8 $\mu$m and a 200-K
object would peak near 14.5 $\mu$m. However, the complex chemistry of these
atmospheres results in a spectral energy distribution far different from a
standard blackbody. Using model atmosphere calculations from Burrows et
al.\ (1997), it is found that the peak flux for brown dwarfs with
1000 K $\la$ T$_{eff}$ $\la$ 200 K is around 4.7 $\mu$m.\footnote{As one
might expect, the exact wavelength of this peak is a function of
T$_{eff}$ and gravity since it is shaped by the wings of strong molecular
absorption bands sculpting it on both sides. Although the
wavelength of the peak can shift either blueward or redward of the quoted 
value by a few tenths of a micron, 4.7 $\mu$m is a reasonable median value 
for the peak over a large range of physical parameters (Burrows, priv.\
comm.).}
The reason is that this 
is one of the few near-infrared wavelengths at which a major absorption 
band is {\it not} found -- H$_2$O, CH$_4$, and H$_2$ provide major sources
of opacity between 1 and 4 $\mu$m, and H$_2$O, CH$_4$, and NH$_3$ provide
more opacity between 5 and 9 $\mu$m. 

The predicted spectral energy distribution for a cool brown dwarf
is shown in Figure 3. Specifically, this is a Burrows et al.\ (1997) model 
for a brown dwarf with T$_{eff} = 500$K and age $\approx$ 1 Gyr. 
Locations and breadths of the NGSS photometric bands are
shown by the shading in the figure. Note the
flux peak between 4.0 and 5.0 $\mu$m and the dominant absorption bands to
either side of this. Cool brown dwarf candidates can thus be 
selected from the lists of NGSS detections by choosing those sources with
extremely red [3.5$\mu$m]-[4.7$\mu$m] colors.

\begin{figure}
\plotfiddle{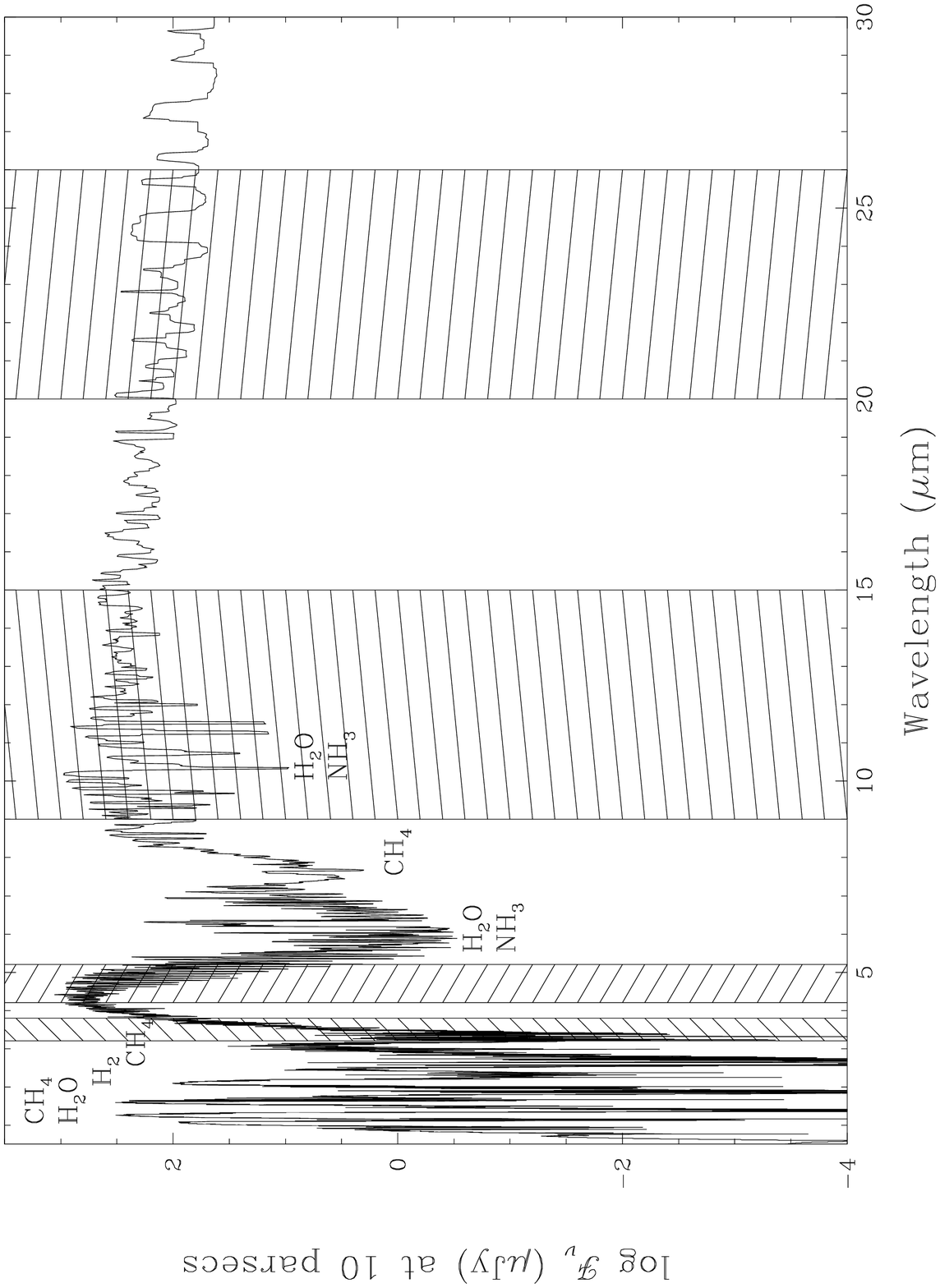}{3.5in}{270}{50}{50}{-195}{300}
\caption{A Burrows et al.\ model of a 500K brown dwarf with an age of $\sim$1
Gyr. Major absorption bands by methane, water, ammonia, and molecular 
hydrogen are shown. The four filter bandpasses of NGSS are shown by the 
shaded regions.}
\end{figure}

Brown dwarfs with T$_{eff} > 800$K, seen at increasingly larger distances from
the Sun, will become non-detections at 12 and 23 $\mu$m before they are
undetectable in the other two NGSS bands. Brown dwarfs with T$_{eff} <
800$K drop out first in the 3.5 and 23 $\mu$m bands. Requiring a
candidate to be detected in two bands (i.e., so there would be a solid 
NGSS-determined color) gives, as a function of brown dwarf T$_{eff}$, the 
distance limits shown in the second column of Table 2. If we resort to
color {\it limits} and require that a source be detected in only a single 
NGSS band
(which is always the 4.7 $\mu$m filter for cool brown dwarfs), we could
in principle probe to the distance
limits given in the third column of Table 2.

\begin{table}
\begin{center}
\caption{Brown Dwarf Detection Limits for NGSS}
\begin{tabular}{ccc}
\tableline
Temperature of& Distance Limit& Distance Limit\\
Brown Dwarf   & Probed, 2-band& Probed, 1-band\\
(K)           & Detections (pc)& Detections (pc)\\
\tableline
1200 & 32& 72\\
 900 & 22& 58\\
 750 & 16& 41\\
 450 &  9& 23\\
 300 &  6& 10\\
 150 &  2&  5\\
\tableline
\tableline
\end{tabular}
\end{center}
\end{table}

To determine the number of cool brown dwarfs that NGSS might uncover, we
are forced to extrapolate based on the rather poorly determined (at present)
field brown dwarf mass function.
Reid et al.\ (1999) made the first such determination
using early discoveries of L and T dwarfs from 2MASS and DENIS. 
They assumed that the mass function is described by a power law
($\Psi(M) = dN/dM \propto M^{\alpha}$) and that the birthrate of brown dwarfs
has been constant over the lifetime of the Galaxy. Assuming various values 
of $\alpha$, they used the evolutionary models of Burrows et al.\ (1997) to
assess the makeup of a complete collection of brown dwarfs in the Solar 
Neighborhood if that value of $\alpha$ were the correct one.
Then using the search criteria employed in finding the earliest L dwarf
discoveries from 2MASS (Kirkpatrick et al.\ 1999), they ran Monte Carlo 
simulations on the collections to derive typical ``observed samples'' of 
brown dwarfs. The set of models
most closely representing the numbers and composition of the actual L dwarfs
found in the 2MASS sample had $1 < \alpha < 2$, with the lower values of the
exponent being preferred. It should be noted that
Chabrier (2002) also finds a value close to $\alpha = 1$ to be the best fit, 
which furthermore
is in agreement with the value for $\alpha$ inferred from brown dwarf
discoveries in young open clusters. 

This work is based on observations of
higher-mass and/or younger brown dwarfs since those
are the only ones to which current surveys are sensitive.
If we extrapolate to lower masses --- into the mass regime 
where NGSS can probe --- we find that there should be at least $\sim$200 
brown dwarfs with $M>10 M_{Jup}$ within 
8 pc of the Sun. A volume limited sample should contain 1.6 times as many 
brown dwarfs with T$_{eff} = 500$ K than 1000K, and ones at 250K should 
outnumber those at 500K by a factor of 2.0. The consequence of these results
on the makeup of the Solar Neighborhood is shown graphically in Figure 4. 

\begin{figure}
\plotone{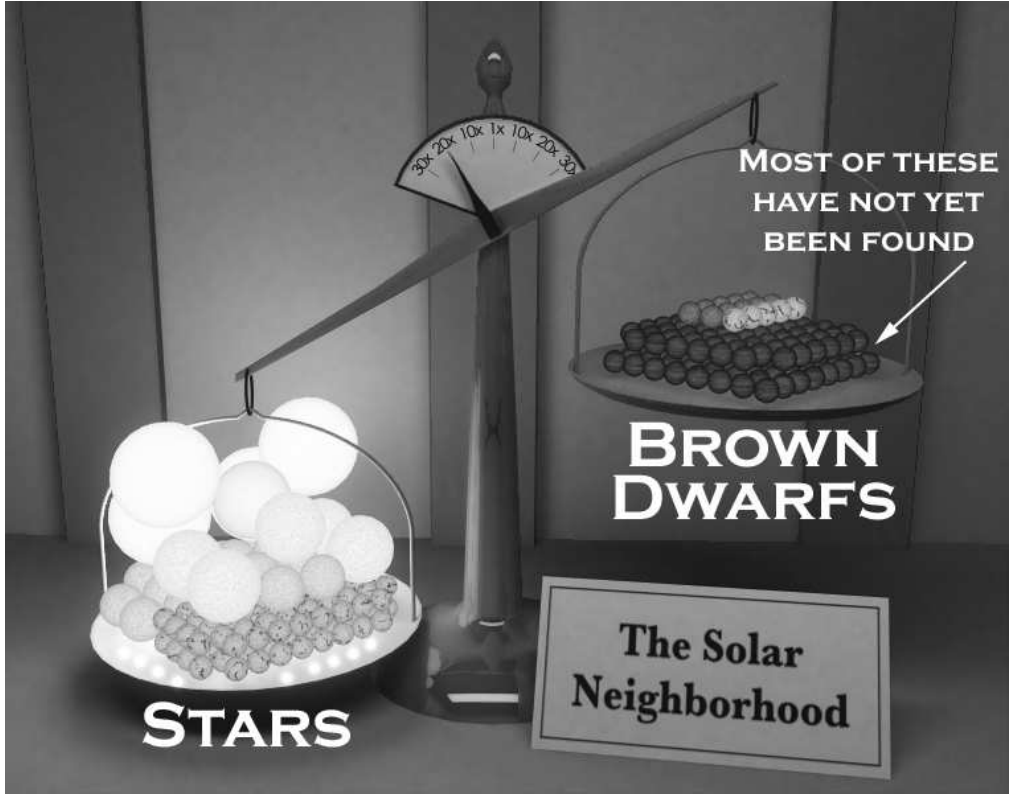}
\caption{The makeup of a typical slice of the Milky Way --- specifically 
the sample of nearby stars with $d < 8$ pc and 
$\delta > -30^{\deg}$. Stars are shown on the left and brown dwarfs (with
mass $>$ 10M$_{Jup}$) on the 
right. Going from the lightest colored, largest stars to the darkest, 
smallest ones,
there are 4 A stars, 1 F star, 5 G stars (one of which is the Sun itself), 
22 K stars, and 87 M stars. Amongst the collection of stars on the left are 
also 9 white dwarfs, shown as small, bright white dots. On the right scale
are brown dwarfs, composed of a few M and L dwarfs along with 
lots of much darker T dwarfs or cooler objects, most of
which are so cool that even 2MASS and SDSS cannot detect them. 
Despite the fact that there are at least as many brown dwarfs as stars, 
brown dwarfs appear to be responsible for only a small fraction of the total
mass, as the scale shows.}
\end{figure}

This implies that the nearby ``stellar'' census is at least
50\% incomplete at present since most of the brown dwarfs are missing. 
Statistically speaking, then, there is a good 
chance that
the nearest brown dwarf lies closer to the Sun than Proxima Centauri at 
$d = 1.3$ pc.
If we again assume $\alpha = 1$ and take as a worst-case scenario 1 
$M_{Jup}$ to be the cutoff mass
for brown dwarf formation, then the median T$_{eff}$ of the present-day
brown dwarf field population should be $\sim$150 K. As Table 2 shows, such
an object is detectable by NGSS in two bands out to $d = 2$ pc.
Superlatives and possible future press releases aside, within 10 pc of the
Sun NGSS can perform a complete census of objects down to at least
35 $M_{Jup}$ 
since the oldest brown dwarfs (at age $\approx$ 10 Gyr) of this mass have 
cooled to 500 K. This mass function will be sampled to even lower 
masses for younger objects. For example, a 500 K brown dwarf at age 100 Myr 
has a mass of 5 $M_{Jup}$. 

\section{Current Status of the NGSS Mission}

Phase-A studies for the current round of MIDEX selections are due in 
October, 2002, and NASA will select the winners
around March, 2003. If successful, NGSS would launch sometime in 2007 or
shortly thereafter. This mission is a necessary precursor to the
Next Generation Space Telescope (NGST) -- hence the similarity in names --
as it will provide science targets and astrometric references for the
NGST, due to launch in 2010.

\vskip 12pt

{\it Acknowledgments:} The author would like to thank the Principal 
Investigator of NGSS, Ned Wright, for proofreading the manuscript and 
offering the graphic used in Figure 1.
The author also thanks the other members of the NGSS Science 
Team (Andrew Blain, Martin
Cohen, Roc Cutri, Peter Eisenhardt, Nick Gautier, Isabelle Hawkins, Tom
Jarrett, Carol Lonsdale, John Mather, Ian McLean, Robert McMillan, Deborah
Padgett, Michael Ressler, Michael Skrutskie, Adam Stanford,
and Russ Walker) for providing their time and support to the project, and
the many members
of the technical and managerial support teams without whom NGSS would never
fly. The
artistic renditions shown in Figures 2 and 4, both of which are publicly
available at the Arhive of M, L, and T Dwarfs at 
\verb"http://spider.ipac.caltech.edu/staff/davy/ARCHIVE/", are the work of 
Robert Hurt and were supported through a grant to HST Proposal \#08563 
provided by the Space Telescope Science Institute, which is operated by the
Association of Universities for Research in Astronomy, Inc., under NASA
contract NAS5-26555. The renderings of
Figure 2 would not have been possible without the assistance of Mark
Marley, who provided guidance on the visible appearance of these objects
using the latest results from model atmosphere calculations. The author
would also like to thank Adam Burrows for providing the spectral model
plotted in Figure 3.


\begin{references}
\reference Burgasser, A. J., et al.\ 2000, ApJL, 531, 57.
\reference Burrows, A., et al.\ 1997, ApJ, 491, 856.
\reference Chabrier, G. 2002, ApJ, 567, 304.
\reference Kirkpatrick, J. D., et al.\ 1999, ApJ, 519, 802.
\reference Reid, I. N., et al.\ 1999, ApJ, 521, 613.
\end{references}
\end{document}